\documentclass{aa}
\usepackage{amssymb}
\usepackage{natbib}
\bibpunct{(}{)}{;}{a}{}{,}
\usepackage[table]{xcolor}
\usepackage{graphicx}
\usepackage{longtable}
\usepackage{graphicx,wasysym}
\usepackage{longtable}
\usepackage{color}

\def\hess{H.E.S.S.\ }
\def\hessns{H.E.S.S.}
\def\msol{M_{\odot}}
\def\gr{$\gamma$-ray\ }
\def\grs{$\gamma$-rays\ }
\def\wco{W_\textrm{CO}}

\begin{document}

\title{On the possible correlation of Galactic VHE source locations and enhancements of the surface density in the Galactic plane}
\author{G.~Pedaletti \inst{1},
E. de O\~{n}a Wilhelmi \inst{1}, 
 \and D. F. Torres \inst{1,2}}

\institute{
Institut de Ci\`encies de l'Espai (IEEC-CSIC),
              Campus UAB,  Torre C5, 2a planta,
              08193 Barcelona, Spain
\and 
Instituci\'o Catalana de Recerca i Estudis Avan\c{c}ats (ICREA)
}

\abstract{}{The association of very-high energy sources with regions of the sky rich in dust and gas has been noticed in the study of individual VHE sources. However, the statistical significance of such correlation for the whole population of TeV detections has not been assessed yet. Here we present a study of the association of VHE sources in the central Galactic region with positions of enhanced material content.}{ We obtain estimates of the material content through two classical tracers: dust emission and intensity of the $^\textrm{12}$CO(1$\rightarrow$0) line. We make use of the recently released all-sky maps of astrophysical foregrounds of the Planck Collaboration and of the extensive existing CO mapping of the Galactic sky. In order to test the correlation, we construct randomized samples of VHE source positions starting from the inner Galactic plane survey sources detected by the \hess array.}{ We find hints of a positive correlation between positions of VHE sources and 
regions rich in molecular material, which in the best of cases reaches the 3.9$\sigma$ level. The latter confidence is however decreased if  variations in  the selection criteria are considered,  what lead us to conclude that a positive correlation cannot be firmly established yet. 
Forthcoming VHE facilities will be needed in order to firmly establish the correlation.}{}


\authorrunning{Pedaletti et al.}
\titlerunning{Galactic VHE sources and surface density enhancements}
\maketitle

\section{Introduction}
Most of the very-high energy (VHE; E$\gtrsim$100 GeV) sources were discovered in the last decade through a scan of the Galactic plane region by the High Energy Stereoscopic System (\hessns). In many cases, multiwavelength studies of the position of VHE emission allowed the classification of the source within a given class of known VHE emitters. Still, many sources are unidentified. It would seem that the Galactic VHE sources are in regions of the sky where the surrounding is rich in molecular material, but no correlation is established up to date.

Such a relation would be expected in a scenario where the VHE emission has an hadronic origin: the accelerated cosmic rays (CRs, protons or heavier nuclei) interact with the surrounding material, leading to VHE emission through the decay of the produced $\pi^0$ \citep{ginzburg1964}. Such emission can be boosted by the presence of an enhancement of target material above the ISM contribution \citep[see e. g., ][]{w28hess}.

Mass enhancements (or simply enhancements of the dust content) can also be associated with regions of high stellar activity, which are in turn associated with known or predicted VHE emitters, such as Pulsar Wind Nebulae (PWN), Supernova remnants (SNR), binaries or regions of massive-star formation. In the case of PWN, the VHE emission needs to take into account also the inverse Compton scattering (ICS) of the accelerated electrons off the thermal radiation associated to the dust \citep[e.g.,][]{JonMartin2012}. 
VHE emission has also been predicted to be produced in regions of massive-star formation, where apart from being obvious sites prone to the appearance of SNRs and other accelerators, the strong winds of the hot OB stars in the clusters form acceleration region at wind interaction zones.  
These regions are rich in molecular material and dust and are hence expected to be bright regions for the tracers that we consider here. The extreme cases in this sense are the ULIRGs and starburst galaxies \citep[e.g.,][]{voelkstarbust,diegostarburst,decea,lackistarburst}, some of which have also been established as high energy emitters both in GeV and TeV \citep{starb_lat,starb_hess}.

Mass content is usually obtained trough well established tracers. The most commonly used is the intensity of the $^\textrm{12}$CO(1$\rightarrow$0) line at 115 GHz (2.6 mm), used as tracer of H2 \citep{dame_co}. However, if the gas is diffuse, the line cannot be excited and this method fails to trace all the mass content of the region. The gas in this phase is usually referred to as ``\textit{dark gas}'' and can be better studied through HE (E$\gtrsim$100 MeV) \gr emission or dust emission \citep{planckdark,planckdust}. In this work, we will use both the CO line and dust emission as tracers.
 
Thanks to the increase of the VHE detections and the richness of data released by the Planck Collaboration \citep[see, e.g., ][]{planckco, planckdust}, it is possible to estimate quantitatively the possible correlation of the population of Galactic VHE sources with enhancements of the mass distribution. We do so next.

\section{The data set and mass estimation}
The mass in a given position in the sky can be expressed as
\begin{equation}\label{eq:mass}
 M=\mu m_\textrm{H} D^2 \Delta\Omega N_{\rm{H}},
\end{equation}
where $\mu=1.4$ is the mean weight per H, $m_\textrm{H}$ is the mass of the H nucleon, $D$ is the distance, $\Delta\Omega$ is the solid angle and $N_{\rm{H}}$ is the hydrogen column density. Therefore one would only need to estimate $N_{\rm{H}}$ if the distance is known.

An extensive archive of CO line (1$\rightarrow$0) at 115 GHz (2.6 mm) intensity $\wco$ is provided in \citet{dame_co} and \citet{newdame_co}. The data are released\footnote{http://www.cfa.harvard.edu/rtdc/CO/} in cubes of radial velocity (in Local Standard of Rest), galactic latitude and longitude. The radial velocity allows the estimation of the distance. From $\wco$ one can estimate the mass of the molecular material as follows: 
\begin{equation}\label{eq:massco}
 M_\textrm{CO}=\mu m_\textrm{H} D^2 \Delta\Omega_\textrm{px} X_\textrm{CO} \displaystyle{\sum\limits_\textrm{px}} \wco
\end{equation}
The suffix relates to the method of estimation, but refers to the total molecular mass. The solid angle is taken out of the summation as it is the same for each pixel (squares of 0.125$^\circ$ per side).

\begin{figure*}[!tbp]
\begin{center}
\includegraphics[width=0.9 \linewidth]{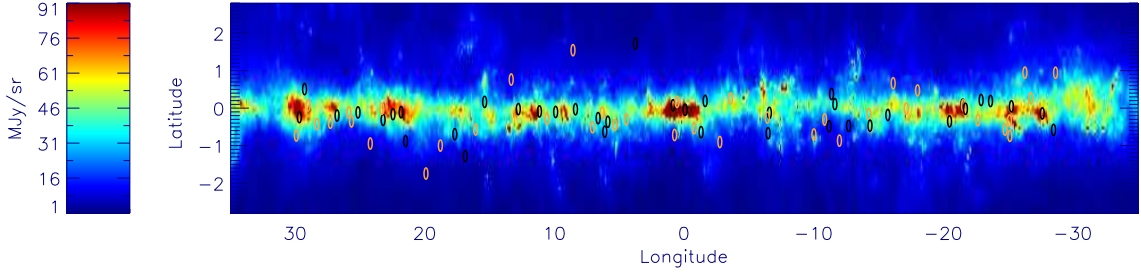}
\caption{Positions overlaid on dust emission map (at 353 GHz, Planck data), cut at 10\% of maximum value. Black circles correspond to GPS sources. Orange circles correspond to a random sample drawn from 2D distribution in Fig. \ref{fig:gpsdistro}. See text for details.}
 \label{fig:gpssimpriors}
\end{center}
\end{figure*}

One can estimate the total mass (hydrogen in any form) from the dust content. From the thermal emission of the dust, we can derive $N_{\rm{H}}$ \citep[see, e. g.][]{roy_orion_dust}. Using Eq. \ref{eq:mass}, the total mass would be:
\begin{equation}\label{eq:massdust}
 M_\textrm{dust}= \Delta\Omega_\textrm{px} D^2 \kappa_{\nu_0}^{-1} \displaystyle{\sum\limits_\textrm{px}}\frac{I_\nu}{B_\nu(T)} \left( \frac{\nu}{\nu_0}\right)^{-\beta},
\end{equation}
where $\kappa_{\nu_0}$ is the mass absorption coefficient (MAC), that contains the knowledge on the emissivity of the dust grains and their composition.
The intensity of the dust emission can be parametrized as follow:
\begin{equation} \label{eq:inu_dust}
 I_\nu = A \left( \frac{\nu}{\nu_0}\right)^\beta B_\nu(T),
\end{equation}
where $B_\nu(T)$ is the Planck function for dust equilibrium temperature $T$, $A$ is the amplitude of the modified blackbody and $\beta$ the dust spectral index. The dust optical depth at frequency $\nu$ is
\begin{equation}
 \tau_\nu=\frac{I_\nu}{B_\nu(T)}=A \left( \frac{\nu}{\nu_0}\right)^\beta.
\end{equation}
The Planck legacy archive\footnote{http://pla.esac.esa.int/pla/aio/planckProducts.html} provides dust parameters as all-sky maps of $\tau$, $\beta$ index and $T$ at 353 GHz (with degeneracy between the last 2 parameters). They were obtained by fitting the Planck data at 353, 545 and 857 GHz together with the IRAS (IRIS) 100 micron data \citep[see][]{planckdust}. One can use all-sky maps for $\tau$ and $\beta$ to calculate the mass. As can be seen from Eq. \ref{eq:inu_dust}, this is equivalent to using the intensity maps, if we restrict ourselves to the emission at 353 GHz (reference frequency for the intensity map available in the archive) as we do in the following. 
The Planck satellite legacy archive also provides $\wco$ in galactic coordinates. This is equivalent to the velocity-integrated Dame map in the full velocity range with the addition of a $<10\%$ contribution from $^\textrm{13}$CO (at 110 GHz, hence in the same HFI channel of the Planck satellite). Here we use only the map from the Dame archive, dubbed \textit{Dame-deep}, that is the composition of several surveys and covers the entire galactic plane at low latitudes ($|b|<5^\circ$). Indeed we will focus only on the central Galactic region. The Dame scan provides the advantage of selecting only one isotopologue thanks to spectroscopic information. Above all, we choose not to use the Planck maps for CO emission to have the two estimate of the mass as independent as possible. 
Using the healpix package\footnote{http://healpix.sourceforge.net/}, the dust maps have been repixelized to match the resolution of the \textit{Dame-deep} maps (1px=0.125$^\circ$). This is slightly larger than the original resolution of the all-sky map of optical depth \citep[5', ][]{planckdust}. The dust emission intensity map at 353 GHz, cut to the inner Galactic region and 10\% of maximum value, is shown in Fig. \ref{fig:gpssimpriors}.

Each estimate of the mass has some unknowns, most noticeably the conversion factors from tracer to mass: the term $X_\textrm{CO}$ for Eq. \ref{eq:massco} and the coefficient $\kappa_{\nu_0}$ in Eq. \ref{eq:massdust}. 
It has to be noted that these are constants, hence their value is not important here as we are more interested in relative mass gradients to define enhancements. Nevertheless, for the scope of this work we will use $X_\textrm{CO} = 1.8 \times 10^{20} \textrm{cm}^{-2} (\textrm{K km/s})^{-1} $ \citep{newdame_co} and $\sigma_{\rm e}(353\mathrm{GHz})=\mu m_{\rm H} \kappa_\nu =2.2 \times 10^{-26}$ cm$^2$ H$^{-1}$. The choice of this particular value is given in the next section. 

A further unknown is the distance of the VHE sources. We therefore will from here onwards refer to the material content through the estimation of the surface density:
\begin{equation}
 \Sigma=\left(\frac{M}{\msol}\right)\left(\frac{A}{\textrm{pc}^{2}}\right),
\end{equation}
where $A$ is the area projected in the sky of the region where the mass $M$ is estimated.
\subsection{Mass absorption coefficient or opacity of the interstellar material}\label{subs:mac}
The MAC value depends on the type of dust considered. 
In \citet{ismplanck} the best estimate of the dust opacity is $\sigma_{\rm e}=\mu m_{\rm H} \kappa_\nu=\tau/N_H = 0.92 \pm 0.05 \times 10^{-25}$  cm$^2$ H$^{-1}$ at 250 $\mu$m(1200GHz), for the ISM. An environment more similar to the mass enhancements studied here is the Orion molecular cloud, where the opacities can be larger by a factor of $\sim 2-4$ \citep[Herschel data from][]{roy_orion_dust}. As they note, much higher opacities than what is obtained for the ISM are not unprecedented, another example
being $\sigma_{\rm e}(1200\mathrm{GHz})=3.8 \times 10^{-25}$ cm$^2$ H$^{-1}$ for a dense molecular
region in the Vela molecular ridge where the column density ranged
between [10 .. 40] $\times10^{21}$ cm$^{-2}$ \citep{Martin2012}.
A study of the impact of dust grain composition is given in \cite{macstudy}, where values as high as $\kappa_{850\mu m}\simeq0.8 \textrm{ m}^2/\textrm{kg}$ are shown. This would correspond to $\sigma_{\rm e}(353\mathrm{GHz})=2 \times 10^{-25}$ cm$^2$ H$^{-1}$, but this is very extreme. 

The  spectral dependence of the opacity can be parametrized as $\sigma_{\rm e}(\nu)/\sigma_{\rm e} (\nu_0) = (\nu/\nu_0)^{\beta}$, with a fiducial frequency $\nu_0 = 1200$~GHz (250~$\mu m$) as in \citet{roy_orion_dust}. Assuming $\beta=1.8$ and considering an average value for the opacity as $\sigma_{\rm e}(1200\mathrm{GHz})=2 \times 10^{-25}$ cm$^2$ H$^{-1}$, the value that we obtain for the frequency used here is $\sigma_{\rm e}(353\mathrm{GHz})=2.2 \times 10^{-26}$ cm$^2$ H$^{-1}$, what we choose for this work.

\section{The inner GPS sources}

\begin{figure}[!hbtp]
\begin{center}
\includegraphics[width=0.9 \linewidth]{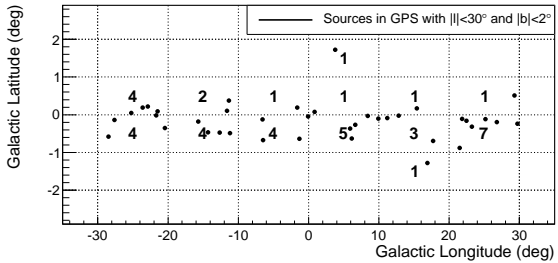}
\caption{Distribution on the sky of the sources in the central galactic plane region. The numbers refer to the abundance of sources in each bin of the grid in which we divided the inner Galactic plane ($\delta l=10^\circ$, $\delta b=1^\circ$).}
 \label{fig:gpsdistro}
\end{center}
\end{figure}

We select the sources in the \hess Galactic plane survey \citep[GPS, as presented in][]{gpsicrc13}\footnote{The online database at http://tevcat.uchicago.edu/ was used for the actual positions. The only exception being HESS J1833-105, that it is not yet present in the webpage.} with $|l|<30^\circ$ and $|b|<2^\circ$. The resulting 39 sources (henceforth referred to as GPS sources) are then binned in space according to their Galactic coordinates (with bins of $\delta l=10^\circ$ and $\delta b=1^\circ$). Their distribution in the sky is shown in Fig. \ref{fig:gpssimpriors} and \ref{fig:gpsdistro}. This is certainly a flux-biased sample, limited by the exposure weighted sensitivity of the \hess array in a certain part of the scan. However the inner Galactic region is the most uniformly covered, providing an almost flat detection threshold for fluxes above 2\% CU (1CU $\sim2.8\times 10^{-11} (E/\textrm{TeV})^{-2.57}$ cm$^{
-2}$ s$^{-1}$ TeV$^{-1}$). 

We proceed to build 10$^5$ fake distributions of TeV sources, by randomizing the locations of each detection yet conserving their number within each spatial bin. One randomized realization is shown in Fig. \ref{fig:gpssimpriors}. In order to be able to construct such a large number of randomized distributions, the binning in space cannot be much smaller than what chosen here or the same random positions would be selected often. The result are proven to be stable with the number of randomization if these are $>10^4$.
The surface density for the random positions is calculated as described above, depending on the mass tracer. We sum the emission in a square of 3x3 pixels, with the central pixel corresponding to the best fit position of the VHE source. As for both maps 1px=0.125$^\circ$, the area considered is similar to the trial extension used in the HESS GPS scan for sources discovery \citep[radius=0.22$^\circ$, see][]{surveyhess}. The error on the source localization is typically smaller than the pixel size considered here. 

In order to quantify the correlation, we calculate the probability of having a certain number of positions of the GPS sources associated with a surface density estimate above a chosen threshold, thus defining the enhancement. The chosen threshold is defined a priori as the surface density for which only 10\% of the positions in the considered inner galaxy have a larger surface density. To be able to compare this threshold directly to the surface density calculated as shown above, we follow the same procedure. The surface density of each point in the galaxy is calculated scanning through each bin of the map and summing the emission of a 3x3 pixel square surrounding said bin. This surface density distribution is oversampled but it is intended to serve only the scope of threshold definition. With the surface density threshold defined, we then construct the distribution function of the randomized samples and its cumulative distribution function, from 
which we can extract the significance of the correlation.

\subsection{Mass enhancement traced by $\wco$ or dust emission}

\begin{figure}[!hbtp]
\begin{center}
\includegraphics[width=1. \linewidth]{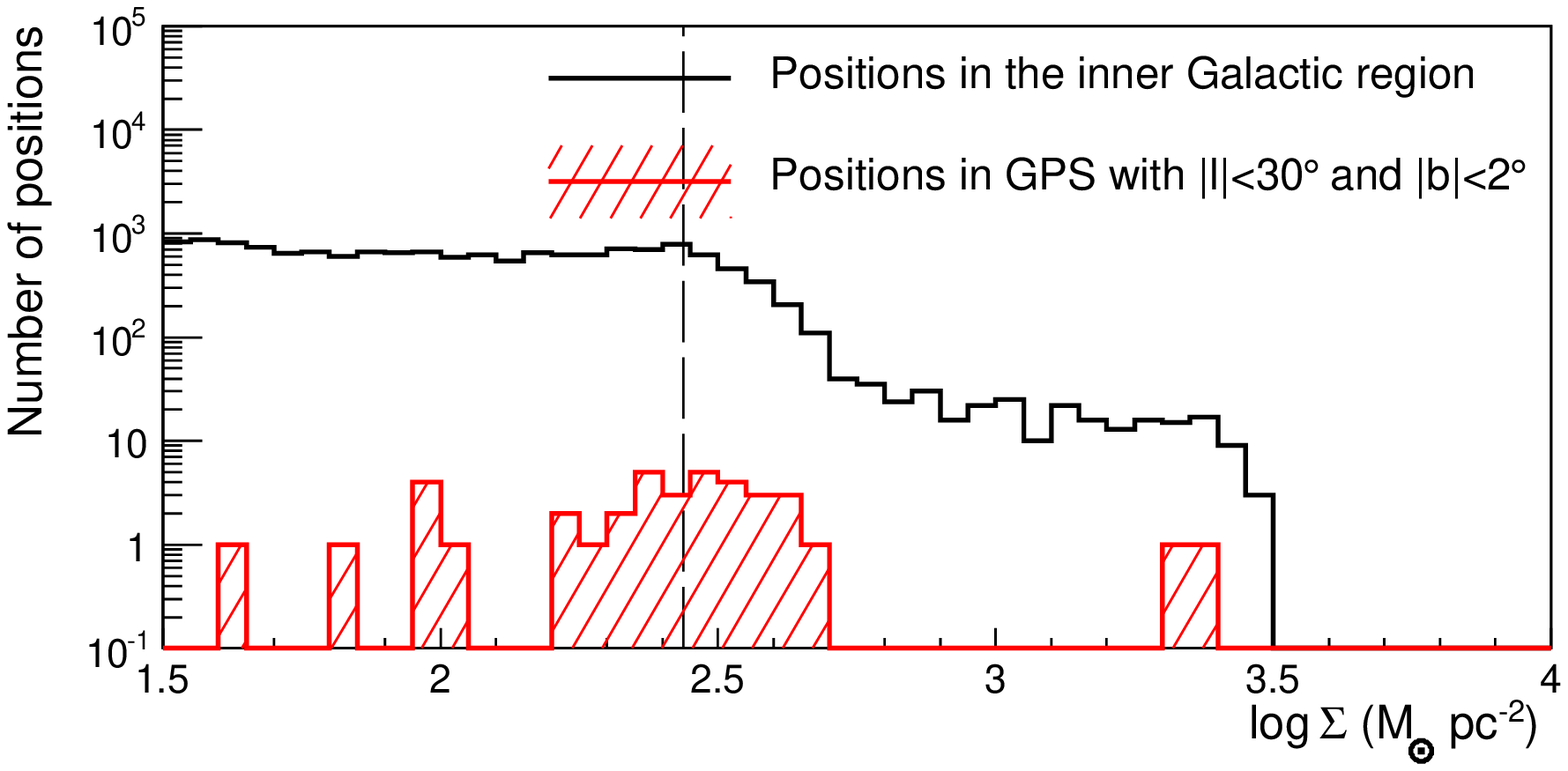}
\includegraphics[width=1. \linewidth]{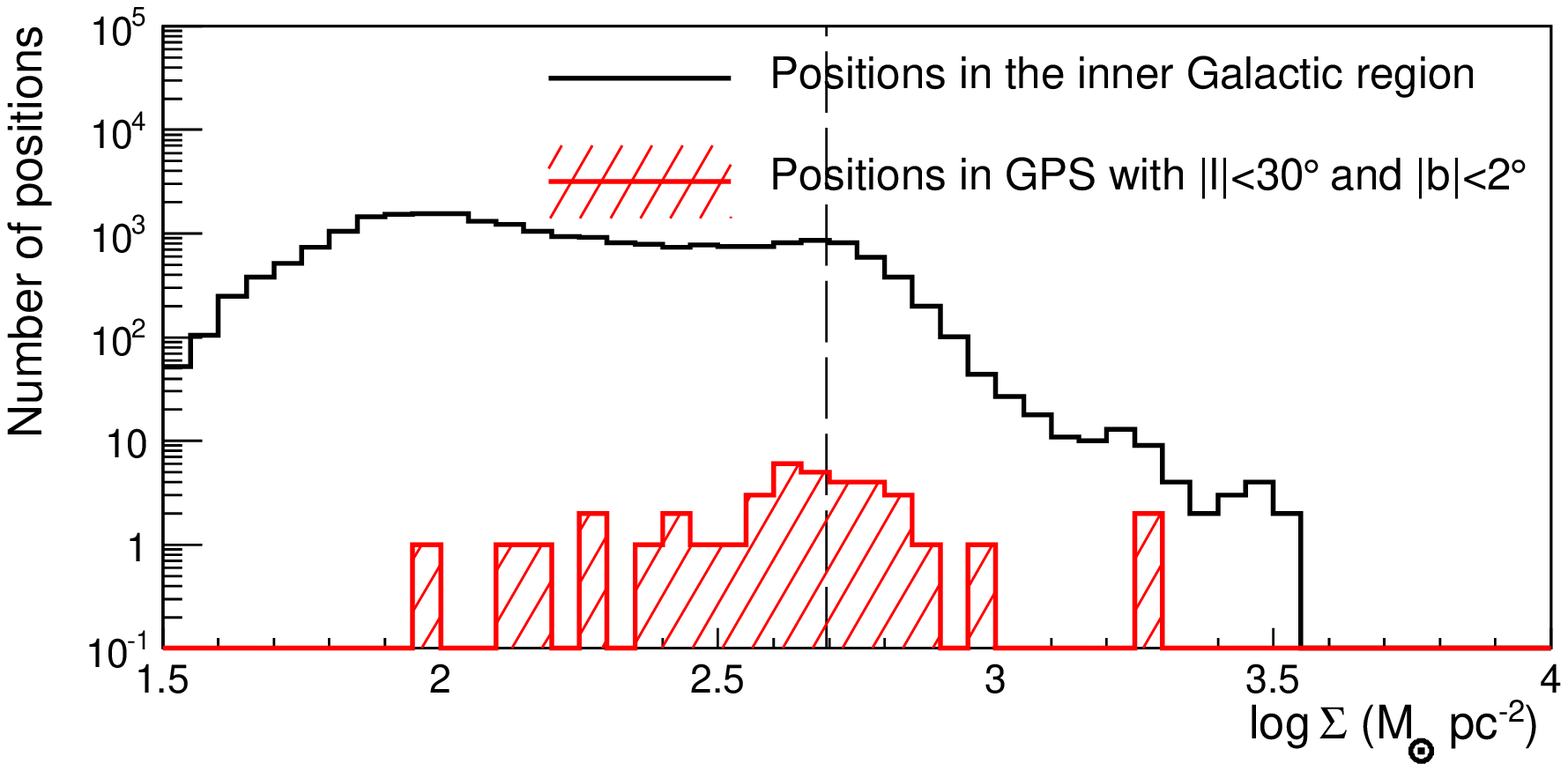}
\caption{Distribution in surface density of the positions in the inner galaxy (black histogram) and in the GPS (red histogram). \textbf{Top, CO:} There are 19 sources above the threshold of $\Sigma_\mathrm{thr,CO}=274\ M_\odot \textrm{pc}^{-2}$ (dashed vertical line). \textbf{Bottom, Dust:} There are 16 sources above the threshold of $\Sigma_\mathrm{thr,dust}=496\ M_\odot \textrm{pc}^{-2}$ (dashed vertical line).}
 \label{fig:massdistro}
\end{center}
\end{figure}

\begin{figure*}[!tbp]
\begin{center}
\includegraphics[width=0.3 \linewidth]{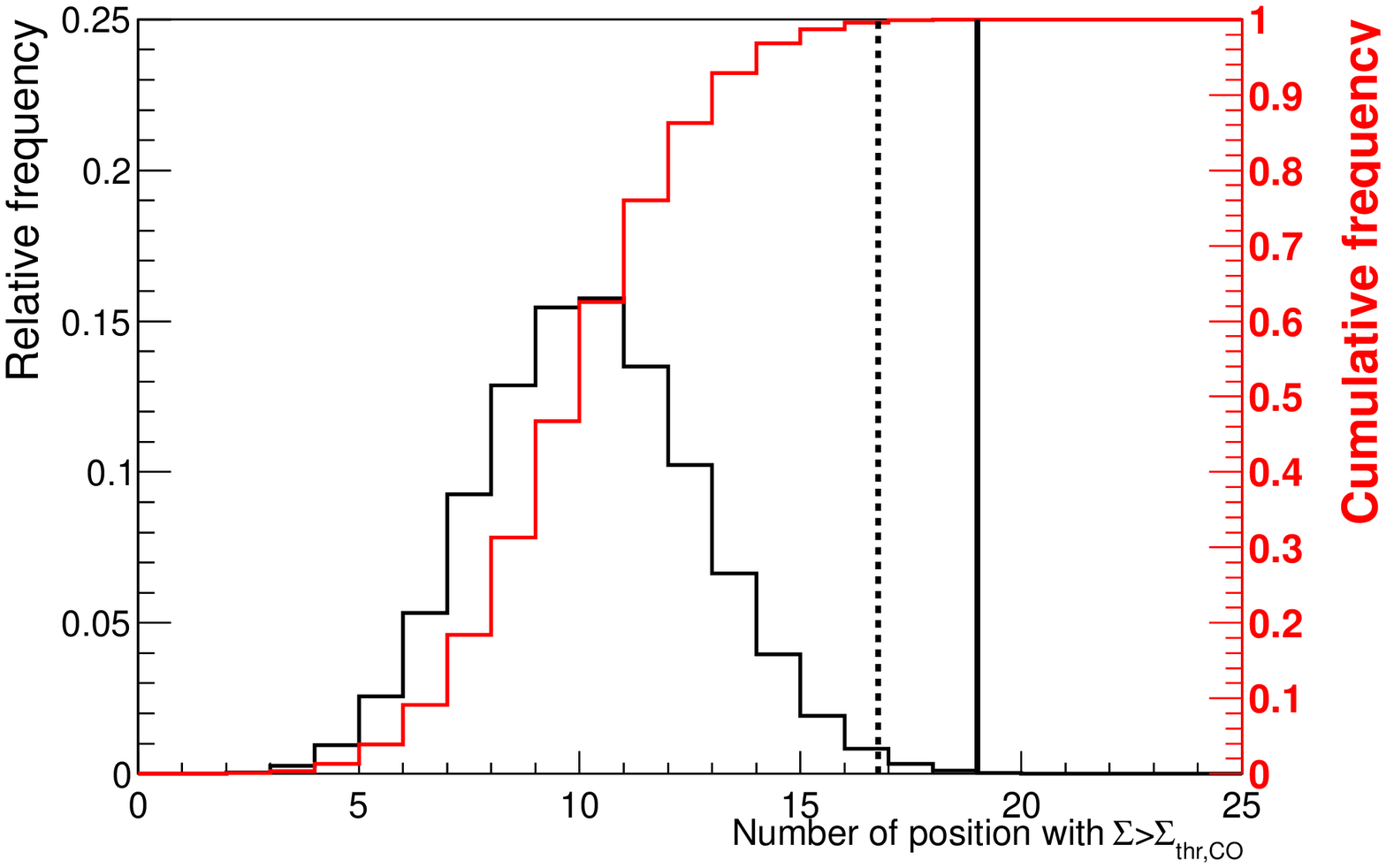}
\includegraphics[width=0.3 \linewidth]{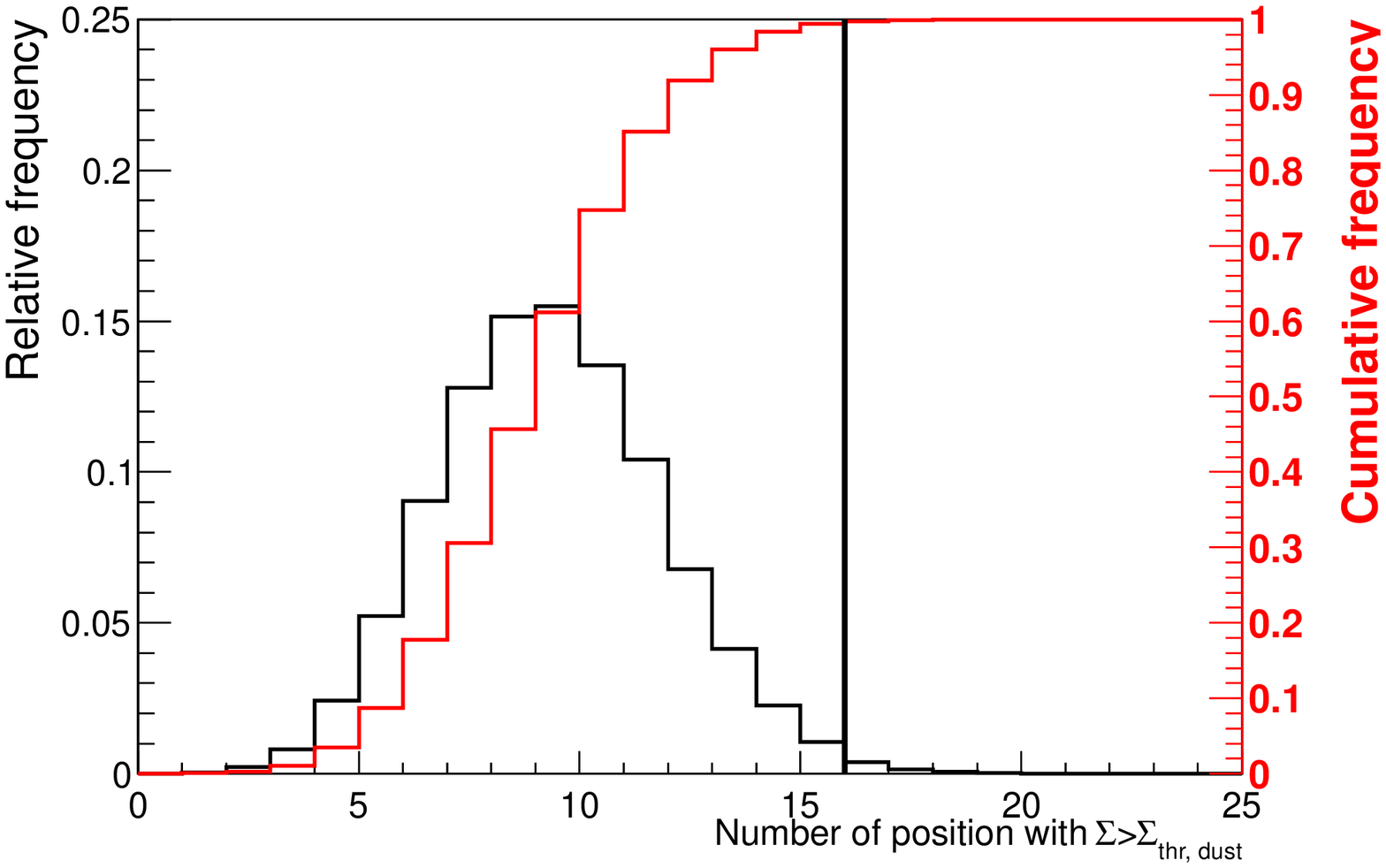}
\includegraphics[width=0.3 \linewidth]{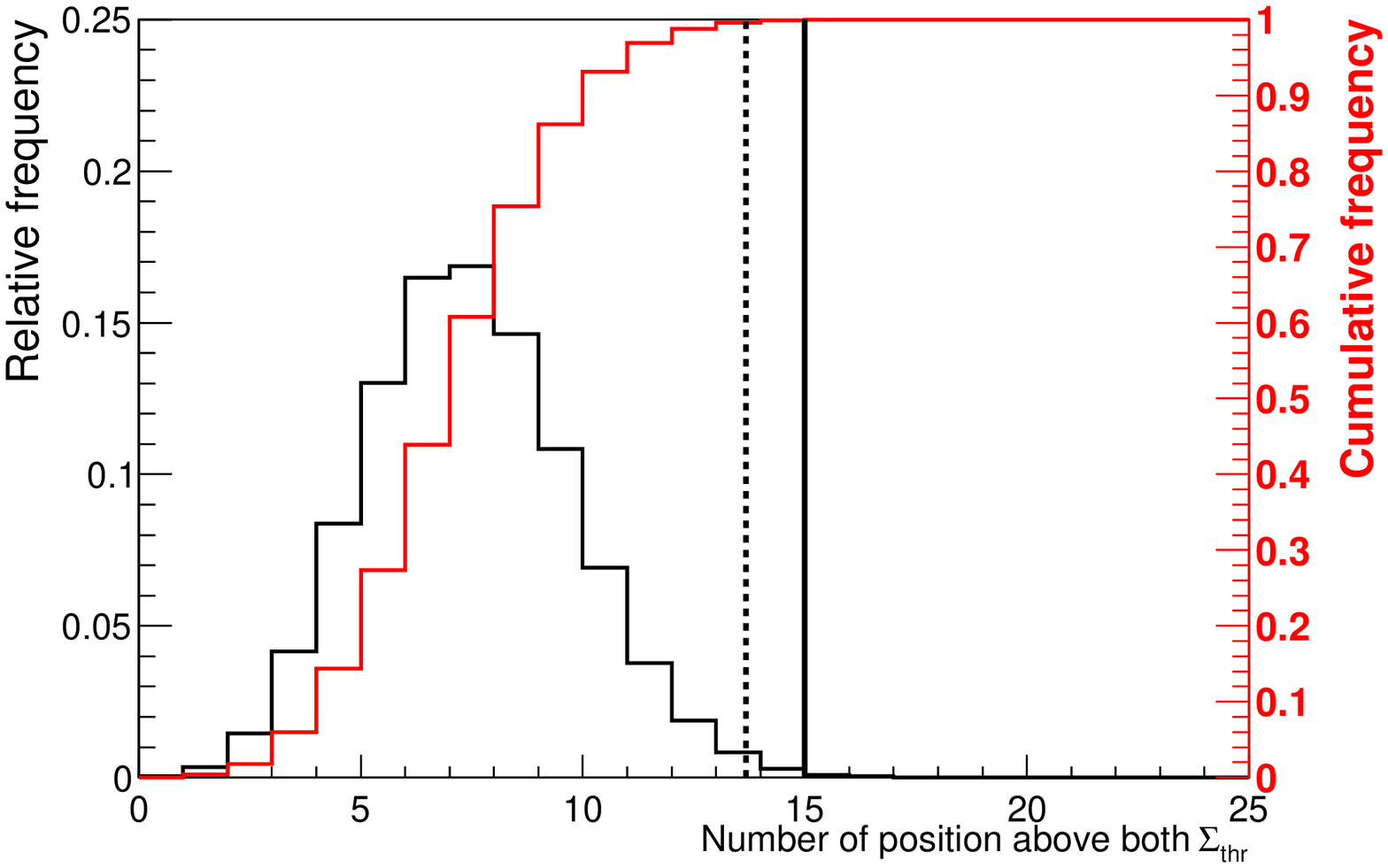}
\caption{Occurrences in randomized sample of positions relative to an integrated surface density $\Sigma>\Sigma_\mathrm{thr,x}$. The dashed
vertical line is for 3$\sigma$. \textbf{Left, CO:} There are 19 sources in the GPS above threshold, that corresponds to 3.9$\sigma$ (solid vertical line). \textbf{Middle, Dust:} There are 16 in the GPS above threshold, that corresponds to 3.1$\sigma$ (solid vertical line). \textbf{Right:} Occurrences in randomized sample where surface density estimates from both tracers are above the respective threshold. There are 15 in the GPS that satisfy simultaneously the two thresholds, and this corresponds to 3.5$\sigma$.}
 \label{fig:abovemass}
\end{center}
\end{figure*}

Following the procedure just explained, we construct the surface density distribution over the entire inner galaxy region estimating the values from the \textit{Dame-deep} $\wco$ map integrated along the line of sight. The surface density threshold is thus defined as $\Sigma_\mathrm{thr,CO}=274 \ M_\odot \textrm{pc}^{-2}$. When applying the threshold to the surface density distribution corresponding to the positions of the GPS sources, we obtain 19 sources in the real GPS sample (see Fig. \ref{fig:massdistro}, top). 
When applying the same threshold to the 10$^5$ randomized set of 39 positions, we obtain the distribution given in Fig. \ref{fig:abovemass} (left). Therefore we can conclude that the significance of correlation between the selected VHE sources and the surface density enhancement that we define is of 3.9$\sigma$ (99.989\%). The significance of the correlation is calculated from the probability extracted from the constructed cumulative distribution, assuming a normal underlying distribution.

The emission of the 2.6 mm CO line might not be tracing all the gas, especially when diffuse. Therefore we repeat the same exercise starting from the dust opacity map in the Planck data repository. 

Considering the surface density distribution in the inner galaxy obtained from the dust emission, a threshold of $\Sigma_\mathrm{thr,dust}=496\ M_\odot \textrm{pc}^{-2}$ is chosen (see Fig. \ref{fig:massdistro}, bottom).
After 10$^5$ realizations of samples of 39 positions, the distribution obtained is the one shown in Fig. \ref{fig:abovemass} (middle). Therefore the GPS sample, with 16 sources showing $\Sigma>\Sigma_\mathrm{thr,dust}$ (see Fig. \ref{fig:massdistro}, bottom), presents a significant correlation at 3.1$\sigma$ significance (99.77\%). 

\subsection{Combined correlation}
As the correlation we are interested about is 
only with the material content, we should compute the probability that the positions of the VHE sources sitting on an enhancement
are the same for both tracers. Out of the 19 positions with $\Sigma>\Sigma_\mathrm{thr,co}$ and 16 with $\Sigma>\Sigma_\mathrm{thr,dust}$, 15 are common to both samples. That is to say, 15 of the GPS sources are associated with a surface density above threshold both in the case of CO and dust. These 15 sources correspond to a significance of 3.5$\sigma$ (99.95\%) when compared to randomized samples. We used here the same randomized sample studied above. For each set of 39 randomized positions, we calculate how many are associated with a surface density above the threshold selected for each tracer.

It appears that the VHE emission correlates more strongly with the dense region well traced by $\Sigma_\textrm{CO}$. Ideally, to prove this, one could search for correlation in smaller subsamples, for example selecting those VHE sources where an hadronic component in the emission is more probable. However, hacking the sample further would incur into significance reduction due to trials and, given the small significances, it is not advisable. 

\subsection{Stability of the result}

The test described in the previous sections
relies on our educated guesses on the construction of the randomized sample and on our definition of surface density enhancement.
We therefore now study the stability of the correlations found.

The $\Sigma_\mathrm{thr}=\Sigma_\mathrm{10}$ was chosen such that only 10\% of the positions in the inner Galaxy presented a larger surface density. If this percentage were to be smaller, this would restrict de facto the study to only the VHE sources in the Galactic center region; if  larger, it would select common values of the surface density distribution (cf. Fig. \ref{fig:massdistro}). We nonetheless have scanned the evolution of significance with percentages ($\Sigma_\mathrm{[5,10,15,20,25]}$, where the numbers in the subscript refers to the percentage). The significance of correlation remains positive ($>3\sigma$) for thresholds of $\Sigma_\mathrm{[10-25]}$ with a peak for $\Sigma_\mathrm{[10,CO]}$ and $\Sigma_\mathrm{[20,dust]}$. 

The initial binning of the sample was chosen in order to allow for a large randomization of positions in a single bin, without incurring in the risk of selecting often the same region of $0.375^\circ\times0.375^\circ$. However, with the initial binning of $\delta b=1^\circ$ (hereafter case A), the latitude distribution of the randomized samples would have a standard deviation of 0.61$^\circ$, slightly larger value than the original sample (0.47$^\circ\pm$0.05$^\circ$). This is not a critical widening, but might impact the final significance assesement if the randomized position were to select on average positions with a smaller surface density than in the respective original GPS position. This is possible, but difficult to quantify given the asymmetrical distribution of the material and the small widening comparable to the size of a single integrated position. We therefore reduce the binning in latitude (from the initial $\delta b=1^\circ$ to $\delta b=0.5^\circ$, hereafter case B), making the width 
of the original and simulated samples (0.51$^\circ$) fully compatible. We reduce the number of randomization to $10^4$. The threshold for the definition of surface density enhancement does not depend on this binning, hence it has the same value as in case A.
The significance of correlation against $\Sigma_\mathrm{10}$ is reduced to $2.4\sigma$ for the CO tracer case and $1.4\sigma$ for the dust tracer case. The significance never exceeds the level of $3\sigma$, but follows the same behaviour described in case A (peak for $\Sigma_\mathrm{[10,CO]}$ and $\Sigma_\mathrm{[20,dust]}$).  

As we are relying on a quantity (the surface density) integrated along the line of sight, we can expect the tests performed to be less sensitive in the regions closer to the Galactic plane where VHE source are more abundant and matter enhancements are more common. We threfore proceed to replicate the test above (case B), but neglecting the region closer to the Galactic plane ($|b|<0.5^\circ$, hereafter case C). We now are focussing on the correspondingly-located 9 VHE sources. Surface density thresholds will change with respect to case A and B and are given in Table \ref{tbl:results} where we summarize the results. We do obtain a positive correlation of $3.5\sigma$ significance for $\Sigma_\mathrm{[5,dust]}$, that however will suffer for trial reduction (the 5 trials of the $\Sigma_\mathrm{thr}$ scan reduce it already to $3\sigma$) and cannot be claimed as a correlation.

We therefore conclude that we cannot firmly claim a correlation neither with CO traced material nor with dust content enhancement, but all the tests performed present hints of it. The final proof of correlation is left for the future, when the forthcoming Cherenkov Telescope Array (CTA) observations will provide us with more sources \citep{survey_cta}.
\begin{table}[!htbp]
 \caption{Significance (S) of correlations with different tracers and $\Sigma_\mathrm{10}$}
\label{tbl:results}
  \centering
\begin{tabular}{cccccc}
 \hline \hline
Case & Tracer & $\Sigma_\mathrm{thr}$ & \#   & Probability & S \\
&& $\msol \textrm{pc}^{-2}$ &&&\\
\hline
A&Dust     &  496             &      16    &   0.9977     &  3.05$\sigma$   \\
A&CO       &  274             &      19    &   0.9999     &  3.89$\sigma$    \\
A&Combined &  --              &      15    &   0.9995     &  3.52$\sigma$    \\
\hline
B&Dust     &  496             &      16    &   0.8366     &  1.39$\sigma$   \\
B&CO       &  274             &      19    &   0.9822     &  2.37$\sigma$    \\
B&Combined &  --              &      15    &   0.9610     &  2.06$\sigma$    \\
\hline
C&Dust     &  266             &      4    &   0.8804     &  1.55$\sigma$   \\
C&CO       &  137             &      4    &   0.9170     &  1.73$\sigma$    \\
C&Combined &  --              &     4    &   0.9663     &  2.12$\sigma$    \\
\hline
\end{tabular}
\end{table}

\section{Implication for diffuse emission.}
\begin{figure*}[!tbp]
\begin{center}
\includegraphics[width=0.9 \linewidth]{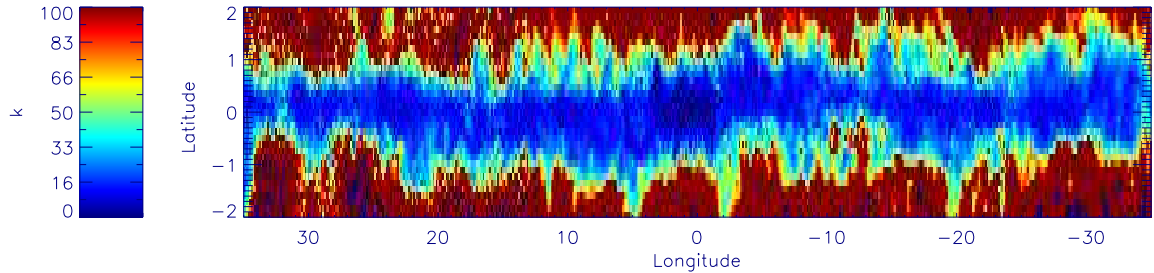}
\caption{Map of the enhancement $\kappa$ over the Galactic CR background necessary to reach detection with 50 hours of CTA observations. Artificially cut to an upper value of $\kappa=100$. All the regions with higher values are at $|b|>1^\circ$.}
 \label{fig:kmap}
\end{center}
\end{figure*}

Let us consider the expected emission from a molecular cloud under the interaction with the cosmic ray sea \citep{aha_passive,gabici_review}:
\begin{equation}\label{eq:passive}
 F(E> E_\gamma) \sim 1 \times 10^{-13}  \kappa E_\gamma^{-1.7} M_\mathrm{5} D ^{-2} \mathrm{cm}^{-2} \mathrm{s}^{-1},
\end{equation}
where $E_\gamma$ is expressed in TeV, the distance $D$ in kpc, the mass $M_\mathrm{5}=10^5 M_\odot$, and $\kappa$ is the enhancement factor of CRs, assumed to be unity for passive clouds (i.e. a case where only the CR background is included) and larger in the presence of a nearby accelerator. For the same level of $\kappa$, the larger the mass content of the target material, the larger the expected flux will be. 
If the mass is calculated through Eq. \ref{eq:mass}, then the dependence on distance is canceled.
Hence it would be possible to translate the $\wco$ or dust emission maps in maps of the minimum expected VHE contribution relative to the Galactic CR background. This minimum level of diffuse emission should be completed with the addition of the expected \grs from the ICS contribution off the Galactic radiation field. The \gr diffuse emission is well studied at GeV energies \citep{egretdiffuse,fermidiffuse}. With the large dataset accumulated with observation by the \hess array, studies of the diffuse component start to be possible also with the current generation of Cherenkov Imaging Arrays, not only in the dense region at the Galactic center \citep{diffusehess}. If the correlation that we have studied between VHE sources and regions of enhanced mass was to be confirmed, it implies that, in 
order to select a diffuse component not contaminated by a nearby source of accelerated particles, the diffuse emission needs to be studied in region with a moderate mass content. However, when removing regions with mass content $\Sigma>\Sigma_\mathrm{thr,x}$, the expected diffuse flux of the remaining regions is quite low. For these regions, assuming $X_\textrm{CO} = 1.8 \times 10^{20} \textrm{cm}^{-2} (\textrm{K km/s})^{-1} $, the flux per pixel obtained through Eq. \ref{eq:passive} and \ref{eq:massco} is one order of magnitude below the CTA point source sensitivity for 50 hours of observation \citep[the pixel area considered here is comparable to a point source for CTA at energies $E\sim 1$ TeV,][]{ctaconcept}.
The necessary factor $\kappa$ of enhancement over the Galactic CR background to assure a detection with CTA ranges from a factor of few up to $\kappa>100$ at the highest galactic latitude investigated here, see Fig. \ref{fig:kmap}. It is important to note that this is an upper limit valid in the assumption that the CR spectrum at every location in the galaxy has the same energy dependence than the CR background. If the spectral dependence changes, the necessary $\kappa$ will be lowered by a factor of even $\sim$10 \citep[assuming a CR spectrum with the same normalization at 1 GeV than the CR background, but with a spectral change from $\Gamma=2.7$ to $\Gamma=2.1$, for details see][]{aha_passive}. For comparison, an enhancement factor of $\kappa\gtrsim10$ is actually found in the W28 region \citep[see][]{gavin_w28}.
The \hess preliminary measurements of the diffuse component in the inner galaxy already show how the minimum level given by p-p interaction is clearly exceeded and VHE sources enhance the CR content above the level detected at Earth. The excess can also be due to unresolved sources. The most promising regions to study the VHE emission due only to the Galactic CR background remain the giant molecular clouds in the Gould Belt, even if they will still prove challenging for CTA due to their large extension \citep{gmc_cta_private}.

\section{Concluding remarks}

We studied the possible correlation of VHE emission and matter content enhancement in the inner Galactic region ($|l|<30^\circ$ and $|b|<2^\circ$). We constructed the probability distribution function of the number of positions in the inner galaxy associated with a mass enhancement. The value of surface density threshold to define enhancement is derived for the inner Galaxy distribution. The results for a representative subsample of the performed tests are summarized in Table \ref{tbl:results}. Independently from the tracer used to estimate the surface density, we conclude that a positive correlation cannot be firmly established yet. 

The number of sources discovered from the \hess galactic plane scan keeps on increasing with larger observation time and refined analysis. This study will benefit from a substantial increase of the number of sources, like the one we can expect from the Galactic plane scan with the forthcoming CTA array.

\begin{acknowledgements}
This work was made possible by the release of the following data to the community: the composite CO survey of the Milky Way at http://www.cfa.harvard.edu/rtdc/CO/ and the all-sky maps for CO and dust at Planck Legacy archive (http://pla.esac.esa.int/pla/aio/planckProducts.html). We also acknowledge the TeVcat archive (http://tevcat.uchicago.edu/).
G.P. acknowledges the 2013 IYAS school organized by the "Ecole doctorale Astronomie \& Astrophysique
d'\^{I}le-de-France".
We acknowledge support from the 
the grants AYA2012-39303, SGR2009-811 and iLINK 2011-0303
 
\end{acknowledgements}

\end{document}